\documentclass[aps,prc,preprint,amsmath,amssymb,showpacs,preprintnumbers,superscriptaddress,nofootinbib]{revtex4-2}
\usepackage{CJK}
\usepackage{graphicx}
\usepackage{dcolumn}
\usepackage{bm}
\usepackage{color}
\usepackage{hyperref}

\allowdisplaybreaks[4]

\begin{document}

\title{Deep-neural-network approach to solving the \emph{ab initio} nuclear structure problem}
\author{Y. L. Yang}
\affiliation{State Key Laboratory of Nuclear Physics and Technology, School of Physics, Peking University, Beijing 100871, China}

\author{P. W. Zhao}
\email{pwzhao@pku.edu.cn}
\affiliation{State Key Laboratory of Nuclear Physics and Technology, School of Physics, Peking University, Beijing 100871, China}

\begin{abstract}
Predicting the structure of quantum many-body systems from the first principles of quantum mechanics is a common challenge in physics, chemistry, and material science.
Deep machine learning has proven to be a powerful tool for solving condensed matter and chemistry problems, while for atomic nuclei it is still quite challenging because of the complicated nucleon-nucleon interactions, which strongly couple the spatial, spin, and isospin degrees of freedom.
By combining essential physics of the nuclear wave functions and the strong expressive power of artificial neural networks, we develop FeynmanNet, a deep-learning variational quantum Monte Carlo approach for \emph{ab initio} nuclear structure.
We show that FeynmanNet can provide very accurate solutions of ground-state energies and wave functions for $^4$He, $^6$Li, and even up to $^{16}$O as emerging from the leading-order and next-to-leading-order Hamiltonians of pionless effective field theory.
Compared to the conventional diffusion Monte Carlo approaches, which suffer from the severe inherent fermion-sign problem, FeynmanNet reaches such a high accuracy in a variational way and scales polynomially with the number of nucleons.
Therefore, it paves the way to a highly accurate and efficient \emph{ab initio} method for predicting nuclear properties based on the realistic interactions between nucleons.
\end{abstract}

\maketitle

\section{Introduction}
Atomic nuclei are self-bound systems consisting of protons and neutrons, which interact with each other via strong interactions.
However, it is a great challenge to describe nuclear structure directly from the fundamental theory of strong interactions, quantum chromodynamics (QCD), due to its non-perturbative nature at the low-energy regime.
The advent of the effective field theory (EFT) paradigm in the early 1990s~\cite{Weinberg1990Phys.Lett.B288,Weinberg1991Nucl.Phys.B3} has opened the way to
linking QCD and the low-energy nuclear structure by establishing nuclear EFTs~\cite{Hammer2020Rev.Mod.Phys.025004}, which are nowadays the main inputs~\cite{Epelbaum2009Rev.Mod.Phys.17731825,Machleidt2011Phys.Rep.175,Gezerlis2013Phys.Rev.Lett.032501,Epelbaum2015Phys.Rev.Lett.122301} to \emph{ab initio} nuclear many-body approaches.
The nuclear EFTs provide the nuclear Hamiltonian with controlled approximations and the corresponding many-nucleon Schr\"odinger equation is then solved with state-of-the-art many-body methods.
Such a combination has achieved a great success in describing many nuclear properties including binding energies and radii~\cite{Wienholtz2013Nature346,Hagen2016Nat.Phys.186, Hu2022Nat.Phys.}, $\beta$ decays~\cite{Gysbers2019Nat.Phys.428}, $\alpha$-$\alpha$ scattering~\cite{Elhatisari2015Nature111}, etc.

Nevertheless, some major challenges remain because the nucleon-nucleon interaction is extremely complex, in contrast to the Coulomb force and/or the van der Waals potential used in atomic and molecular physics.
It contains a strong tensor component involving both the spin and isospin of the nucleons and also significant spin-orbit forces, inducing strong coupling between the spin-isospin and spatial degrees of freedom~\cite{Ring2004}.
These features lead to complex nuclear many-body phenomena, whose description requires a consistent treatment of both short-range (or high-momentum) and long-range (or low-momentum) correlations.
Among the variety of nuclear many-body methods, quantum Monte Carlo (QMC) methods~\cite{Carlson2015Rev.Mod.Phys.1067} based upon Feynman path integrals formulated in the continuum have proven to be quite valuable for these problems.
They are able to deal with a wide range of momentum components of the interaction and, thus, can accommodate ``bare'' potentials derived within nuclear EFTs.
However, the QMC methods are presently limited to either light nuclei with up to $A = 12$ nucleons~\cite{Pudliner1995Phys.Rev.Lett.4396,Wiringa2002Phys.Rev.Lett.182501,Lovato2013Phys.Rev.Lett.092501,Piarulli2018Phys.Rev.Lett.052503} or larger systems but with simplified nuclear Hamiltonians~\cite{Gandolfi2007Phys.Rev.Lett.022507,Lonardoni2018Phys.Rev.Lett.122502}.
This is mainly because of the infamous fermion-sign problem~\cite{Troyer2005Phys.Rev.Lett.170201}, which leads to an exponential increasing ratio of error to signal with the number of nucleons.
Therefore, an accurate and polynomial scaling solution is highly desired to extend the QMC calculations to medium-mass nuclei.

Machine learning has provided the opportunity for a polynomial scaling solution of quantum many-body problems, especially for many-electron systems~\cite{Carleo2019Rev.Mod.Phys.45002}.
It is motivated by the fact that artificial neural networks (ANNs) can compactly represent complex high-dimensional functions and, thus, should be able to provide efficient means for representing the wave function of quantum many-body states.
A variational representation of ANN-based many-body quantum states has been originally introduced for prototypical spin lattice systems~\cite{Carleo2017Science602}, and then generalized to several quantum systems in continuous space~\cite{Ruggeri2018Phys.Rev.Lett.205302,Han2019J.Comput.Phys.108929}.
Recently, deep neural networks trained within variational Monte Carlo (VMC) have been further developed to tackle \emph{ab initio} chemistry problems~\cite{Pfau2020Phys.Rev.Research033429,Hermann2020Nat.Chem.891897,Choo2020NatureCommunications2368,Scherbela2022Nat.Comp.Sci.331341}.

For \emph{ab initio} nuclear structure, due to the complexity of the nucleon-nucleon interaction, the application of machine-learning approaches to nuclear many-body problems is still in its infancy.
They are often split into two main categories, supervised and unsupervised~\cite{Boehnlein2022Rev.Mod.Phys.031003}.
Here, the many-nucleon Schr\"odinger equation is solved directly with unsupervised learning.
The first attempt was given to solve deuteron, a two-body bound state, in momentum space~\cite{Keeble2020Phys.Lett.B135743}.
Subsequently, an ANN quantum state ansatz defined by the product of a Jastrow factor and a Slater determinant was introduced to solve nuclei with up to $A=6$ nucleons in coordinate space within the VMC method~\cite{Adams2021Phys.Rev.Lett.022502,Gnech2021FewBodySystems7}.
It outperforms the routinely employed ansatz based on two- and three-body Jastrow functions, while there are still significant deviations from the numerically exact results for three- and four-body nuclei, mainly caused by the incorrect nodal surface of the single Slater determinant.\footnote{
Because of the fermionic nature of the nuclear many-body wave function, it must have a nodal surface where its value takes zero. For the ANN ansatz in Ref.~\cite{Adams2021Phys.Rev.Lett.022502}, the nodal surface is mainly determined by the single trial Slater determinant, which is fixed during the variation process and, thus, usually incorrect.
}
The incorrect nodal surface can be largely improved with an augmented Slater determinant involving hidden nucleonic degrees of freedom~\cite{Lovato2022Phys.Rev.Research043178}, but the nuclear Hamiltonian is limited to contain central forces only, thereby preventing the application to realistic nuclear structure problems, which depend crucially on the tensor and spin-orbit forces~\cite{Wiringa2002Phys.Rev.Lett.182501}.

A key development improving the nodal surface in the present work is the consideration of a many-body backflow transformation, which was originally proposed by Feynman and Cohen for liquid helium~\cite{Feynman1956Phys.Rev.1189}.
While the traditional backflow does not reach a very high accuracy, a series of recent works showed that representing the backflow with a neural network is a powerful generalization~\cite{Luo2019Phys.Rev.Lett.226401} and can greatly improve the accuracy in solving many-electron problems~\cite{Pfau2020Phys.Rev.Research033429,Hermann2020Nat.Chem.891897}.

In this work, we develop a novel deep-learning QMC approach for nuclear many-body problems, FeynmanNet, which includes multiple Slater determinants and backflow transformation based on powerful deep-neural-network representations encompassing both continuous spatial and discrete spin-isospin degrees of freedom for nucleons.
In particular, to incorporate many-body correlations induced by the tensor and spin-orbit forces, the deep neural networks are designed to represent complex-valued nuclear wave functions.
Moreover, physics related to low-energy nuclear structure including the major shell structure and the point symmetries is explicitly encoded in the neural-network architecture, and it makes the obtained FeynmanNet not only highly accurate, but also robust and efficient in the training process.
We demonstrate the high performance of FeynmanNet by benchmarking our results against the hyperspherical-harmonics (HH) method for $^{4}$He and $^{6}$Li and the auxiliary-field diffusion Monte Carlo (AFDMC) approach for $^{16}$O.
Considering that FeynmanNet scales polynomially with the number of nucleons, the present work opens the way to highly accurate \emph{ab initio} studies of medium-mass nuclei with quantum Monte Carlo approaches.

\section{Architecture}
At the core of our approach is a deep-learning architecture, dubbed FeynmanNet, designed for a compact representation of the nuclear wave function.
Due to the strong tensor and spin-orbit interactions among nucleons, it is essential to explicitly write the nuclear wave function to be complex-valued,
\begin{equation}\label{Eq.psi_cplx}
	\Psi(\bm x_1,\ldots,\bm x_A)=\Psi^{(\mathrm{R})}(\bm x_1,\ldots,\bm x_A)+\mathrm{i}\Psi^{(\mathrm{I})}(\bm x_1,\ldots,\bm x_A),
\end{equation}
where $\bm x_i=(\bar{\bm r}_i,s_i,t_i)$ are the single-nucleon variables, including the intrinsic spatial coordinates $\bar{\bm r}_i=\bm r_i-\bm r_\mathrm{c.m.}$ with $\bm r_\mathrm{c.m.}$ being the position of the center of mass, the spin $s_i=\pm1/2$, and the isospin $t_i=\pm1/2$.
The introduction of the intrinsic spatial coordinates $\bar{\bm r}_i$ assures the translational invariance of the wave function and avoids the spurious center-of-mass motions~\cite{Carlson2015Rev.Mod.Phys.1067}.

Both the real and imaginary parts of the wave function are constructed by considering Jastrow correlations and multiple Slater determinants consisting of
backflow transformed orbitals,
\begin{equation}\label{Eq.psi}
  \Psi^{(\alpha)}(\bm x_1,\ldots,\bm x_A)=\mathrm{e}^{\mathcal{U}^{(\alpha)}(\bm x_1,\ldots,\bm x_A)}\sum_{n=1}^{N_\mathrm{det}}w_n^{(\alpha)}\det[\mathbf{f}^{(\alpha,n)}(\bm x_1,\ldots,\bm x_A)],\quad \alpha=\mathrm{R},\mathrm{I}.
\end{equation}
Here, $\mathcal{U}^{(\alpha)}$ are the permutation-invariant Jastrow factors, $N_\mathrm{det}$ the number of Slater determinants, and $w_n^{(\alpha)}$ the weight of the corresponding Slater determinant.
The weights $w_n^{(\alpha)}$ are determined variationally during the training process.

\begin{figure*}[!htpb]
    \centering
	\includegraphics[width=0.9\textwidth]{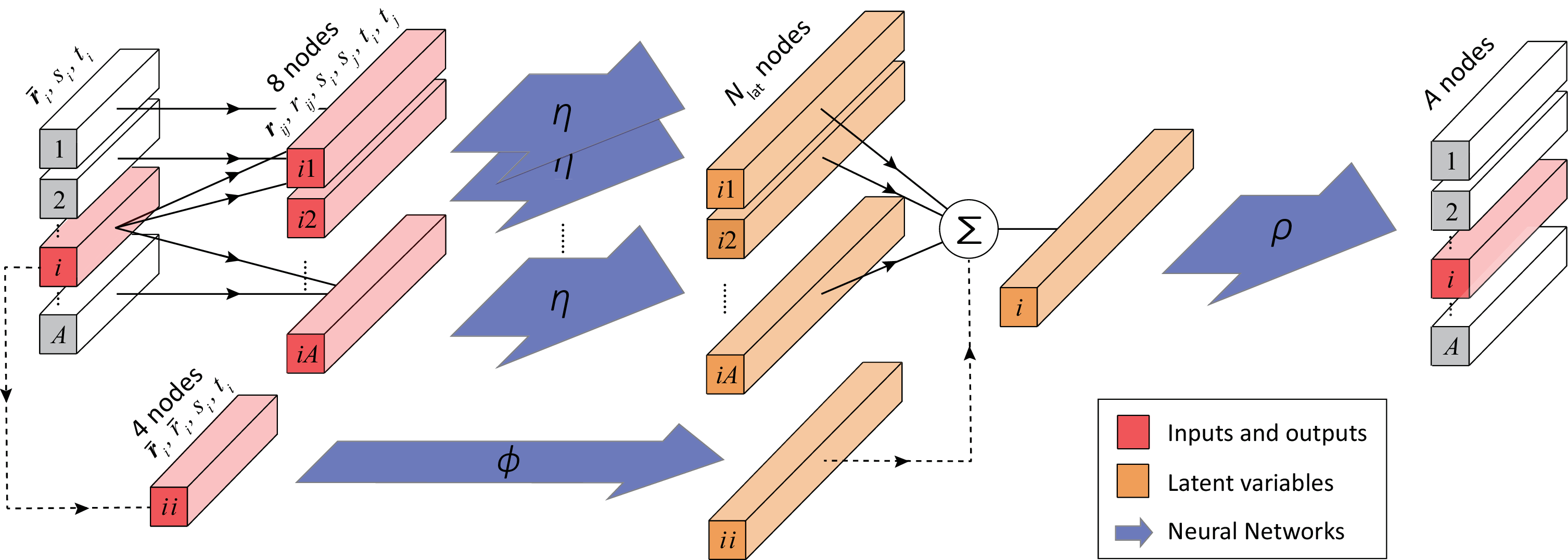}
	\caption{(Color online). Architecture of a backflow neural network in FeynmanNet.
    The input single-nucleon variables of $A$ nucleons are transformed row by row to the $A\times A$ Slater matrix elements consisting of the backflow transformed orbitals. $\phi$, $\eta$, and $\rho$, feed-forward neural networks; $N_\mathrm{lat}$, number of latent variables for each row.
	}\label{fig1}
\end{figure*}

Following the basic idea of the backflow transformation~\cite{Feynman1956Phys.Rev.1189}, the single-nucleon orbitals of the $i$th nucleon in the determinant depend not only on its own variables $\bm x_i$, but also on the variables of all other nucleons in an exchangeable way.
Specifically, as the architecture illustrated in Fig.~\ref{fig1}, the matrix elements of $\mathbf{f}^{(\alpha,n)}$ are represented row by row with neural networks,
\begin{equation}\label{Eq.psi.form}
	f^{(\alpha,n)}_{i\mu}(\bm x_1,\ldots,\bm x_A)=
	\rho_\mu^{(\alpha,n)}\left(\phi^{(\alpha,n)}(\bm x_{ii})+\sum_{j\neq i}\eta^{(\alpha,n)}(\bm x_{ij})\mathrm{e}^{-r_{ij}^2/R^2}\right),
\end{equation}
where $\mu= 1, \ldots, A$ is the index of the orbitals for the $i$th nucleon.

First, the single-nucleon variables $\bm x_i=(\bar{\bm r}_i,s_i,t_i)$ for the $i$th nucleon are combined with those of all other nucleons $j\,(j\neq i)$ to form pairwise inputs $\bm x_{ij}=(\bm r_{ij}, r_{ij}, s_i, s_j, t_i, t_j)$ with $\bm r_{ij}=\bar{\bm r}_i-\bar{\bm r}_j$ and $r_{ij}=|\bm r_{ij}|$.
In principle, the distances $r_{ij}$ are redundant inputs.
However, it was found that inputting $r_{ij}$ could improve the performance of the neural-network wave functions in both electronic~\cite{Pfau2020Phys.Rev.Research033429} and nuclear~\cite{Yang2022Phys.Lett.B137587} systems.
This should be due to the fact that the inclusion of distances $r_{ij}$ respects the rotational invariance of the ground state.

Then, the pair-wise inputs are successively mapped into $N_{\rm lat}$ latent variables via a feed-forward neural network $\eta^{(\alpha,n)}$.
The summation of these latent variables over $j$ assures the permutation invariance for nucleons other than the $i$th nucleon.
The Gaussian function $\mathrm{e}^{-r_{ij}^2/R^2}$ in Eq.~(\ref{Eq.psi.form}), with $R$ as a hyperparameter characterizing the range of nuclear force, is adopted to reduce the correlations of two nucleons which are outside the interacting range.
For the case of $j=i$, the pairwise inputs should be reduced to $\bm x_{ii}=(\bar{\bm r}_i, \bar{r}_i,  s_i, t_i)$, and they are mapped into $N_{\rm lat}$ latent variables via another feed-forward neural network $\phi^{(\alpha,n)}$.
The summation of these latent variables over all nucleon pairs are then input to a new feed-forward neural network $\bm\rho^{(\alpha,n)}$ with $A$ outputs, providing the $A$ matrix elements for the $i$th row.
The designed architecture ensures the antisymmetry of the nuclear wave function, because one can exchange two nucleons by swapping two rows of the matrix $\mathbf{f}^{(\alpha,n)}$, the determinant of which then changes its sign.

The Jastrow factors $\mathcal{U}^{(\alpha)}$ in Eq.~(\ref{Eq.psi}) are represented by neural networks similar to the ones adopted in the previous work~\cite{Gnech2021FewBodySystems7} based on the Deep Sets architecture~\cite{Zaheer2018arXiv1703.06114,Wagstaff2019arXiv.1901.09006}.
The pair-wise inputs of each pair of nucleons $(i,j)$ are mapped separately into a latent-space representation, and a summation over all pairs is then applied to enforce permutation invariance,
\begin{equation}
	\mathcal{U}^{(\alpha)}(\bm x_1,\bm x_2,\ldots,\bm x_A)=\rho^{(\alpha,\mathcal{U})}\left(\sum_{i\neq j}\phi^{(\alpha,\mathcal{U})}(\bm r_{ij}, r_{ij}, s_i, s_j, t_i, t_j)\right).
\end{equation}
Here, $\phi^{(\alpha, \mathcal{U})}$ and $\rho^{(\alpha,\mathcal{U})}$ are feed-forward neural networks.

In the present work, all the feed-forward neural networks, namely $\eta$, $\phi$, and $\rho$, are comprised of one fully-connected hidden layer with 16 nodes. Each of them translates mathematically into the following mapping from the inputs to the outputs,
\begin{equation}
  \bm x_{\rm out}=\sigma\left(\mathbf{W}[\sigma(\mathbf{V}\bm x_\mathrm{in}+\bm a)]+\bm b\right).
\end{equation}
In the above equation, $\mathbf{W}$, $\mathbf{V}$, $\bm a$, and $\bm b$ are the weights and biases of the network, which serve as variational parameters of the wave function.
The activation function $\sigma$ is taken to be the \textit{Softplus} function~\cite{Dugas2001472478}.
The number of latent variables $N_{\rm lat}$, i.e., the output dimensions of $\eta$ and $\phi$, as well as the input dimension of $\rho$, is taken to be 16.

Besides the essential antisymmetry, we also encode other physical knowledge about the nuclear wave function into FeynmanNet, and this significantly strengthens the expressive power of the network and accelerates the training process.
First, the major shell structure of nuclei is embedded in each Slater determinant in Eq.~(\ref{Eq.psi}) by replacing the matrix elements with
$f_{i\mu}^{(\alpha,n)}(\bm x_1,\ldots,\bm x_A)\cdot \varphi_{\mu}(\bm x_i)$, where $\varphi_{\mu}(\bm x_i)$ takes the form
\begin{equation}\label{Eq.shell}
  \varphi_\mu(\bm x_i)=\sum_{k=1}^{N_f}w_{\mu k}\tilde{\varphi}_k(\bm x_i).
\end{equation}
Here, $\tilde{\varphi}_k(\bm x_i),\ k=1,2,\ldots,N_f$, is a set of single-particle shell model orbitals within a closed major shell $(nl)$, and $w_{\mu k}$ the expansion coefficients determined variationally during the training process.
The shell model orbitals are of the form
\begin{equation}\label{Eq.sp.orbits}
	\tilde{\varphi}_{k}(\bm x_i)=R_{nl}(\bar{r}_i)Y_{ll_z}(\hat{\bm r}_i)\chi_{st}(s_{i},t_{i}),
\end{equation}
where $R_{nl}$ are radial functions of a harmonic oscillator
\begin{equation}
	R_{1s}(r)=\mathrm{e}^{-r^2/2b^2},\quad R_{1p}(r)=r\mathrm{e}^{-r^2/2b^2},\quad \ldots,
\end{equation}
$Y_{ll_z}$ the spherical harmonics, and $\chi_{st}\in\{\uparrow_n,\downarrow_n,\uparrow_p,\downarrow_p\}$ the spinors in the spin-isospin space.

In this work, we take the oscillator length $b^2=10\ \mathrm{fm}^2$, and using a different value should not affect FeynmanNet after training.
The harmonic oscillator orbitals up to the $1s$ shell are adopted for $^{4}$He, and $1p$ shell for $^6$Li and $^{16}$O.

Moreover, FeynmanNet explicitly preserves the total isospin projection on the $z$ axis $T_z$ and the parity $\pi$ by writing the nuclear wave function as
\begin{equation}\label{Eq.sym}
  \Psi^{\pi}_{A,Z}(\bm x_1,\ldots,\bm x_A)=\delta_{T_z,\frac{A}{2}-Z}(1+\pi\hat{\mathcal{P}})\Psi(\bm x_1,\ldots,\bm x_A),
\end{equation}
where $A$ and $Z$ are, respectively, the mass and proton numbers of nuclei, and $\hat{\mathcal{P}}$ denotes the operator of space inversion.
For even-even nuclei, the time-reversal symmetry of the wave function is additionally imposed by multiplying $(1+\hat{\mathcal{T}})$ with $\hat{\mathcal{T}}$ being the time-reversal operator.

\section{Training details}
FeynmanNet is trained with the VMC approach by minimizing the energy expectation
\begin{equation}\label{Eq.VMC}
	E[\Psi]=\frac{\langle\Psi\lvert\hat{H}\rvert\Psi\rangle}{\langle\Psi\lvert\Psi\rangle}.
\end{equation}
The stochastic reconfiguration method~\cite{Sorella2005Phys.Rev.B241103}, closely related to the natural gradient descent method~\cite{Amari1998NeuralComputation251} in unsupervised learning, is employed in the training progress to minimize the energy iteratively.
During the training, the parameters at iteration $t$ are updated as
\begin{equation}\label{Eq.SR}
	\bm{p}_{t+1}=\bm{p}_t-\gamma [\mathrm{Re}(\mathbf{S}_t)]^{-1}\bm g_t,
\end{equation}
where $\gamma=5\times10^{-4}$ is the learning rate, $\bm{g}$ is the gradient of the energy $\partial_{\bm p} E$,
\begin{equation}
	g_a=2\mathrm{Re}\left(\frac{\langle \partial_{p_a} \Psi\lvert\hat{H}\rvert\Psi\rangle}{\langle\Psi\lvert\Psi\rangle}-E\frac{\langle \partial_{p_a} \Psi\lvert\Psi\rangle}{\langle\Psi\lvert\Psi\rangle}\right),
\end{equation}
and $\mathbf{S}$ is a precondition matrix
\begin{equation}\label{Eq.SR.matrix}
	\mathrm{S}_{ab}=\frac{\langle \partial_{p_a} \Psi\lvert\partial_{p_b}\Psi\rangle}{\langle\Psi\lvert\Psi\rangle}-\frac{\langle \partial_{p_a} \Psi\lvert\Psi\rangle}{\langle\Psi\lvert \Psi\rangle}\frac{\langle\Psi\lvert\partial_{p_b}\Psi\rangle}{\langle\Psi\lvert\Psi\rangle}.
\end{equation}
Only the real part of $\mathbf{S}$ is employed because the parameters in the neural networks are real-valued.
Moreover, to achieve a robust and efficient training process, the matrix elements $\mathrm{S}_{ab}$ associated with the mixed derivatives with respect to the parameters in the neural network $\eta$ and the parameters in other networks are neglected.

In practice, the precondition matrix $\mathbf{S}$ could be ill-conditioned, namely, with very small eigenvalues, and its inversion could lead to numerical instability.
Therefore, the precondition matrix is regularized by $\mathbf{S}\rightarrow\mathbf{S}+\epsilon\mathrm{diag}(\sqrt{\bm v_t}+10^{-8})$ with the regularization parameter $\epsilon=10^{-3}$ and $\bm v_t=\beta \bm v_{t-1}+(1-\beta)\bm g^2_t$~\cite{Lovato2022Phys.Rev.Research043178}.
Here, $\bm v_t$ accumulates the exponentially-decaying averages of the squared gradients and $\beta$ is the exponential decay factor taken to be 0.9.
In addition, a constraint on the Fubini-Study distance between the wave functions of two adjacent iterations
\begin{equation}
  d_\mathrm{FS}[\Psi(\bm p^{t+1}), \Psi(\bm p^{t})]=\arccos\sqrt{\frac{\lvert\langle\Psi(\bm p^{t+1})\lvert\Psi(\bm p^{t})\rangle\rvert^2}{\langle\Psi(\bm p^{t})\lvert\Psi(\bm p^{t})\rangle\langle\Psi(\bm p^{t+1})\lvert\Psi(\bm p^{t+1})\rangle}}<d_\mathrm{max},
\end{equation}
is employed to prevent accidental large changes of the parameters that might lead to instability.
The limit $d_\mathrm{max}$ is initially set to be 0.1 and lowered to 0.05 when the iteration nearly converges.

At each iteration, a large set of configuration samples $(\bm x_1^{(n)},\ldots,\bm x^{(n)}_A)$ with $\ n=1,\ldots,N$ is generated following the probability distribution $\lvert \Psi\rvert ^2$ by the standard Metropolis Monte Carlo sampling~\cite{Metropolis1953TheJournalofChemicalPhysics10871092}.
Then, the energy expectation $E$, gradient $\bm g$, and precondition matrix $\mathbf{S}$ are evaluated on these samples as in conventional variational Monte Carlo approaches,
\begin{equation}
  \begin{split}
    E&=\mathrm{Re}\langle E^{(n)}\rangle,\\
    g_a&=2\mathrm{Re}\left[\langle O^{*(n)}_aE^{(n)}\rangle-E\langle O^{*(n)}_a\rangle\right],\\
    \mathrm{S}_{ab}&=\langle O^{*(n)}_aO^{(n)}_b\rangle-\langle O^{*(n)}_a\rangle\langle O^{(n)}_b\rangle.
  \end{split}
\end{equation}
Here, the brackets denote the averages over the $N$ configuration samples and
\begin{equation}
  E^{(n)}=\frac{\hat{H}\Psi(\bm x_1^{(n)},\ldots,\bm x^{(n)}_A)}{\Psi(\bm x_1^{(n)},\ldots,\bm x^{(n)}_A)},\quad O_a^{(n)}=\frac{\partial_{p_a}\Psi(\bm x_1^{(n)},\ldots,\bm x^{(n)}_A)}{\Psi(\bm x_1^{(n)},\ldots,\bm x^{(n)}_A)}.
\end{equation}
The derivatives of the wave function with respect to either the spatial coordinates or the neural-network parameters are calculated based on the automatic differentiation framework of TensorFlow~\cite{Abadi2015}.

\section{Nuclear Hamiltonian}
The nuclear Hamiltonian adopted in this work is derived within the pionless EFT, which is based on the tenet that the typical momenta of nucleons in nuclei are much smaller than the pion mass~\cite{Hammer2020Rev.Mod.Phys.025004}.
The nuclear Hamiltonian reads
\begin{equation}\label{Eq.H}
	\hat{H}=\sum_{i=1}^A\frac{-\nabla_i^2}{2m_N}+\sum_{i<j}v_{ij}+\sum_{i<j<k}V_{ijk},
\end{equation}
where $m_N$ is the nucleon mass, $A$ the number of nucleons, $v_{ij}$ the nucleon-nucleon ($NN$) interaction, and $V_{ijk}$ the three-nucleon ($3N$) interaction.
The $NN$ interactions consists of an electromagnetic (EM) term and charge-independent (CI) contact terms at leading order (LO) and additionally charge-dependent (CD) contact terms at next-to-leading-order (NLO),
\begin{equation}
    \begin{split}
    	v_\mathrm{LO}&=v^\mathrm{EM}+v_\mathrm{LO}^\mathrm{CI},\\
    	v_\mathrm{NLO}&=v^\mathrm{EM}+v_\mathrm{LO}^\mathrm{CI}+v_\mathrm{NLO}^\mathrm{CI}+v_\mathrm{NLO}^\mathrm{CD}.
    \end{split}
\end{equation}
The Coulomb repulsion between finite-size (rather than point-like) protons is considered for $v^\mathrm{EM}$~\cite{Wiringa1995Phys.Rev.C38}.
The contact interactions are regularized by Gaussian cutoff functions~\cite{Schiavilla2021Phys.Rev.C054003}, and can be conveniently expressed in terms of radial functions multiplying spin and isospin operators.
The CI contact terms take the form
\begin{equation}\label{Eq.NN.CI}
	v_\mathrm{LO}^\mathrm{CI}(\boldsymbol r_{ij})=\sum_{p=1}^4v_\mathrm{LO}^p(r_{ij})\mathcal{O}_{ij}^p,\quad v_\mathrm{NLO}^\mathrm{CI}(\bm r_{ij})=\sum_{p=1}^8v_\mathrm{NLO}^p(r_{ij})\mathcal{O}_{ij}^p,
\end{equation}
with $\bm r_{ij}=\bm r_i-\bm r_j$, $r_{ij}=\lvert\bm r_{ij}\rvert$, and $\mathcal{O}_{ij}^{p=1,2,\ldots,8}=\mathbf{1},\ \boldsymbol\tau_i\cdot\boldsymbol\tau_j,\ \boldsymbol\sigma_i\cdot\boldsymbol\sigma_j,\ \boldsymbol\sigma_i\cdot\boldsymbol\sigma_j\boldsymbol\tau_i\cdot\boldsymbol\tau_j,\ S_{ij},\ S_{ij}\boldsymbol\tau_i\cdot\boldsymbol\tau_j,\ \boldsymbol L \cdot\boldsymbol S,\ \boldsymbol L \cdot\boldsymbol S\boldsymbol\tau_i\cdot\boldsymbol\tau_j$.
Here, $\bm\sigma_i$ ($\bm\tau_i$) are the Pauli spin (isospin) matrices of the $i$th nucleon, and $S_{ij}=3\boldsymbol\sigma_i\cdot\hat{\boldsymbol{r}}_{ij}\boldsymbol{\sigma}_i\cdot\hat{\boldsymbol{r}}_{ij}-\boldsymbol\sigma_i\cdot\boldsymbol\sigma_j$,
$\boldsymbol{L}=-\frac{\mathrm{i}}{2}\bm r_{ij}\times(\boldsymbol\nabla_i-\boldsymbol\nabla_j)$, and
$\boldsymbol S=\frac{1}{2}(\boldsymbol\sigma_i+\boldsymbol\sigma_j)$ are the tensor operator, the relative angular momentum, and the total spin of a pair of nucleons $(i,j)$, respectively.
Only central forces ($p=1$-4) are present at LO, while the tensor and spin-orbit forces ($p=5$-8) appear at NLO.
The CD contact term at NLO takes the form
\begin{equation}\label{Eq.NN.CD}
	v_\mathrm{NLO}^\mathrm{CD}(\bm r_{ij})=v_\mathrm{NLO}^T(r_{ij})T_{ij},
\end{equation}
with $T_{ij}=3\tau_{iz}\tau_{jz}-\boldsymbol\tau_i\cdot\boldsymbol\tau_j$ being the isotensor operator of the nucleon pair $(i,j)$.
The specific expressions of the radial functions in Eqs.~(\ref{Eq.NN.CI}) and (\ref{Eq.NN.CD}) can be found in Ref.~\cite{Schiavilla2021Phys.Rev.C054003}.

The regularized $3N$ contact interaction reads
\begin{equation}\label{Eq.3N}
	V_{ijk}(r_{ij},r_{jk},r_{ki})=\frac{c_E}{f_\pi^4\Lambda_\chi}\frac{(\hbar c)^6}{\pi^3R_3^6}\sum_\mathrm{cyc}\mathrm{e}^{-(r_{ij}^2+r_{jk}^2)/R_3^2},
\end{equation}
where $\Lambda_\chi=1\ \mathrm{GeV}$, $f_\pi=92.4\ \mathrm{MeV}$ is the pion decay constant, $c_E$ is a three-nucleon low-energy constant (LEC), and $\sum_\mathrm{cyc}$ stands for the cyclic permutation of $i,j,k$.

The LECs in the nuclear Hamiltonian are adjusted to the experimental $NN$ scattering data and $^3$H binding energy~\cite{Schiavilla2021Phys.Rev.C054003}, and we use the optimal set (model ``o") with $R_3=1.0\ \mathrm{fm}$ at LO and $R_3=2.0\ \mathrm{fm}$ at NLO given in Ref.~\cite{Schiavilla2021Phys.Rev.C054003} that was proved to yield reasonably well ground-state energies for several light- and medium-mass nuclei~\cite{Schiavilla2021Phys.Rev.C054003}.

The range of the adopted $NN$ is typically 2 fm, so we use this value for $R$ in the backflow neural network in Eq.~(\ref{Eq.psi.form}).
In addition, for the LO Hamiltonian, only the real part of the FeynmanNet wave function is needed as the tensor and spin-orbit forces are not present [Eq.~(\ref{Eq.NN.CI})].
\section{Results and discussion}
\begin{figure*}[!htpb]
    \centering
	\includegraphics[width=0.9\textwidth]{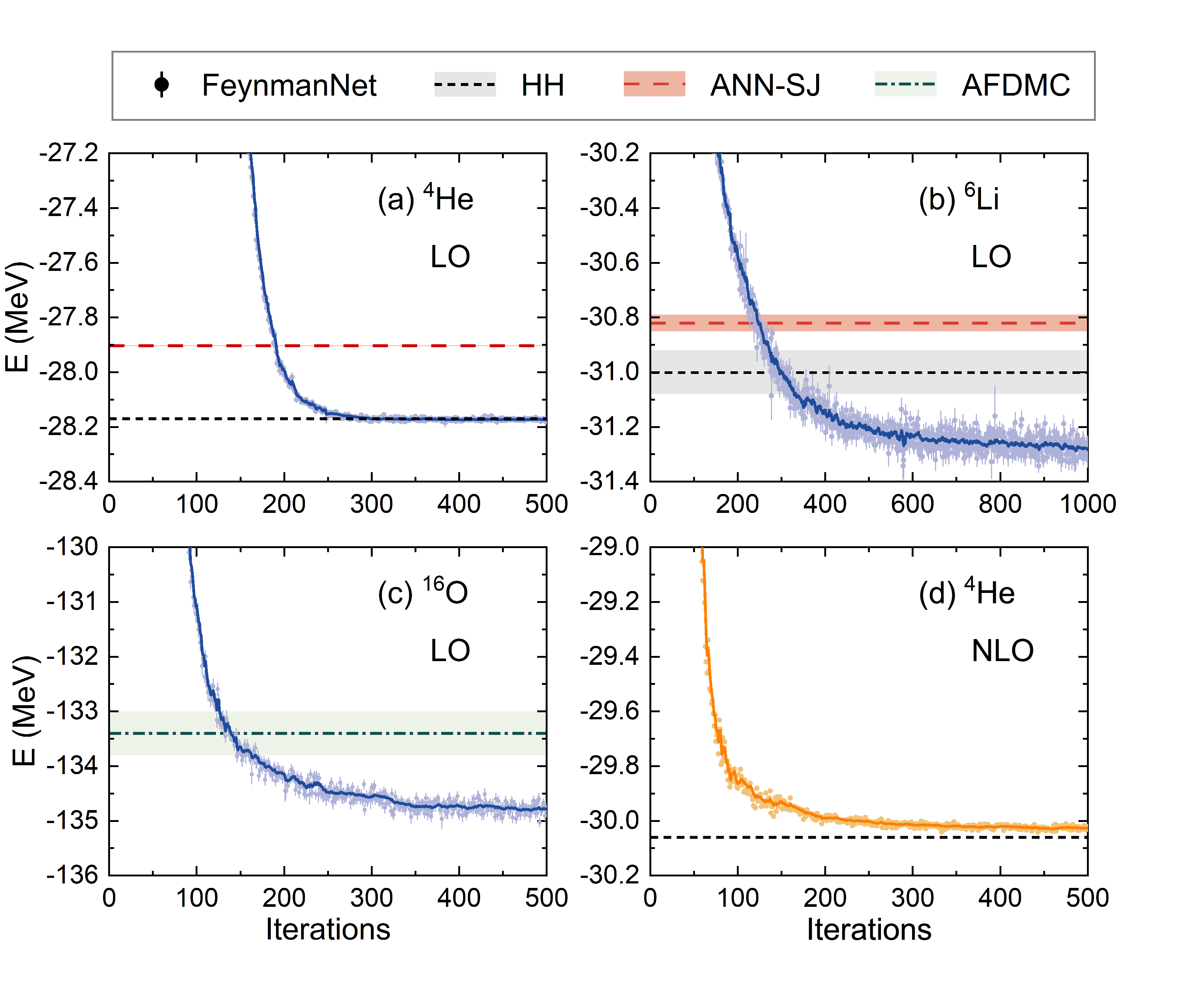}
	\caption{(Color online). Performance of FeynmanNet on the $^4$He, $^6$Li, and $^{16}$O ground states.
	(a) The $^4$He energy, calculated with the pionless effective field theory Hamiltonian at leading order (LO),  as a function of the iterations in the training progress of FeynmanNet. The statistical errors of the energies from the Metropolis Monte Carlo sampling are shown by error bars. The solid line is obtained by applying exponential moving average to the energies. The ground-state energies given by the artificial neural network with Slater-Jastrow (ANN-SJ) ansatz and the hypershperical-harmonics (HH) method~\cite{Gnech2021FewBodySystems7} are displayed for comparison.
	(b) Same as (a) but for $^6$Li. The ANN-SJ and HH results are displayed, with shadow areas indicating the corresponding statistical and extrapolation errors, respectively.
    (c) Same as (a) but for $^{16}$O. The ground-state energy provided by the auxiliary-field diffusion Monte Carlo (AFDMC) method~\cite{Schiavilla2021Phys.Rev.C054003} is displayed with shadow areas indicating the statistical error.
	(d) Same as (a) but with the Hamiltonian at next-to-leading-order (NLO).
    The results of FeynmanNet with the LO and NLO Hamiltonians are shown in blue and orange, respectively.}
\label{fig2}
\end{figure*}

Figure~\ref{fig2} depicts the performance of FeynmanNet by taking $^4$He, $^6$Li and $^{16}$O as examples.
The FeynmanNet results here are obtained using $N_\mathrm{det}=4$ determinants.
For $^4$He and $^6$Li, the obtained ground-state energies are compared with the results given by the previous ANN Slater-Jastrow (ANN-SJ) ansatz and the HH method~\cite{Gnech2021FewBodySystems7}.
The former works only for the LO Hamiltonian, while the latter is valid for both LO and NLO Hamiltonians and, more importantly, is numerically exact for $s$-shell nuclei, e.g., $^4$He.
For the $^4$He ground-state energy at LO (Fig.~\ref{fig2}a), FeynmanNet provides lower energy than the ANN-SJ ansatz after training for only about 200 iterations, and the final result is also consistent with the numerically exact HH value.
This indicates that FeynmanNet outperforms the ANN-SJ ansatz by introducing the multiple determinants and backflow transformation, which improves the nodal surface in both continuous spatial and discrete spin-isospin spaces.
Note that the extra energy given by FeynmanNet grows dramatically for heavier nuclei, e.g., about 7 MeV for $^{16}$O (see the MD-SJ result with $N_{\rm det}=1$ in Fig.~\ref{fig4}c).

The experimental value of $^4$He ground-state energy is $-$28.30 MeV, slightly lower than the HH value, $-$28.17 MeV~\cite{Gnech2021FewBodySystems7}.
However, since the HH method provides a numerically exact solution of the Schr\"odinger equation for $^4$He, this discrepancy originates from the model of the nuclear force, which is out of the scope for this work.

Unlike the $s$-shell nucleus $^4$He, the $p$-shell nucleus $^6$Li is strongly clustered in an $\alpha$ particle and a deuteron, and such a cluster structure brings additional complexity in the calculations.
As a result, the HH result for $^6$Li is not as accurate as that for $^4$He~\cite{Gnech2020Phys.Rev.C014001}.
FeynmanNet converges to the lowest ground-state energies for $^6$Li in comparison with the ANN-SJ and HH results (Fig.~\ref{fig2}b), which are respectively higher by about 500 keV and 300 keV than the FeynmanNet energy.

The expressive power of FeynmanNet is further highlighted for a larger system $^{16}$O.
Such a system is too large for the HH method, so we benchmark our results with the AFDMC approach~\cite{Schiavilla2021Phys.Rev.C054003}.
One can see that the energy given by FeynmanNet is lower than the AFDMC energy by more than 1 MeV (Fig.~\ref{fig2}c).
Note that the AFDMC calculations adopt the constrained-path approximation to mitigate the fermion-sign problem in imaginary-time propagations, and therefore could not solve the ground state exactly~\cite{Wiringa2000Phys.Rev.C014001}.
In contrast, the strong expressive power of FeynmanNet allows a variational approach to reach accurate solutions without performing imaginary-time propagations.

Moreover, the calculation of $^4$He with the NLO Hamiltonian demonstrates the ability of FeynmanNet to deal with tensor and spin-orbit forces (Fig.~\ref{fig2}d).
The NLO Hamiltonian has lower symmetries than the LO one. At LO, the spatial and spin angular momenta, namely $\bm L$ and $\bm S$, are respectively conserved in addition to the total angular momentum $\bm J$.
However, they are broken by the tensor and spin-orbit forces introduced at NLO.
Despite these difficulties, it is remarkable that the number of iterations for convergence of FeynmanNet at NLO is similar to that at LO.
FeynmanNet reaches an accuracy of $\simeq 50$ keV after training for only 200 iterations, and the energy obtained after 500 iterations is consistent with the HH value within 30 keV.

\begin{figure*}[!htpb]
    \centering
	\includegraphics[width=0.9\textwidth]{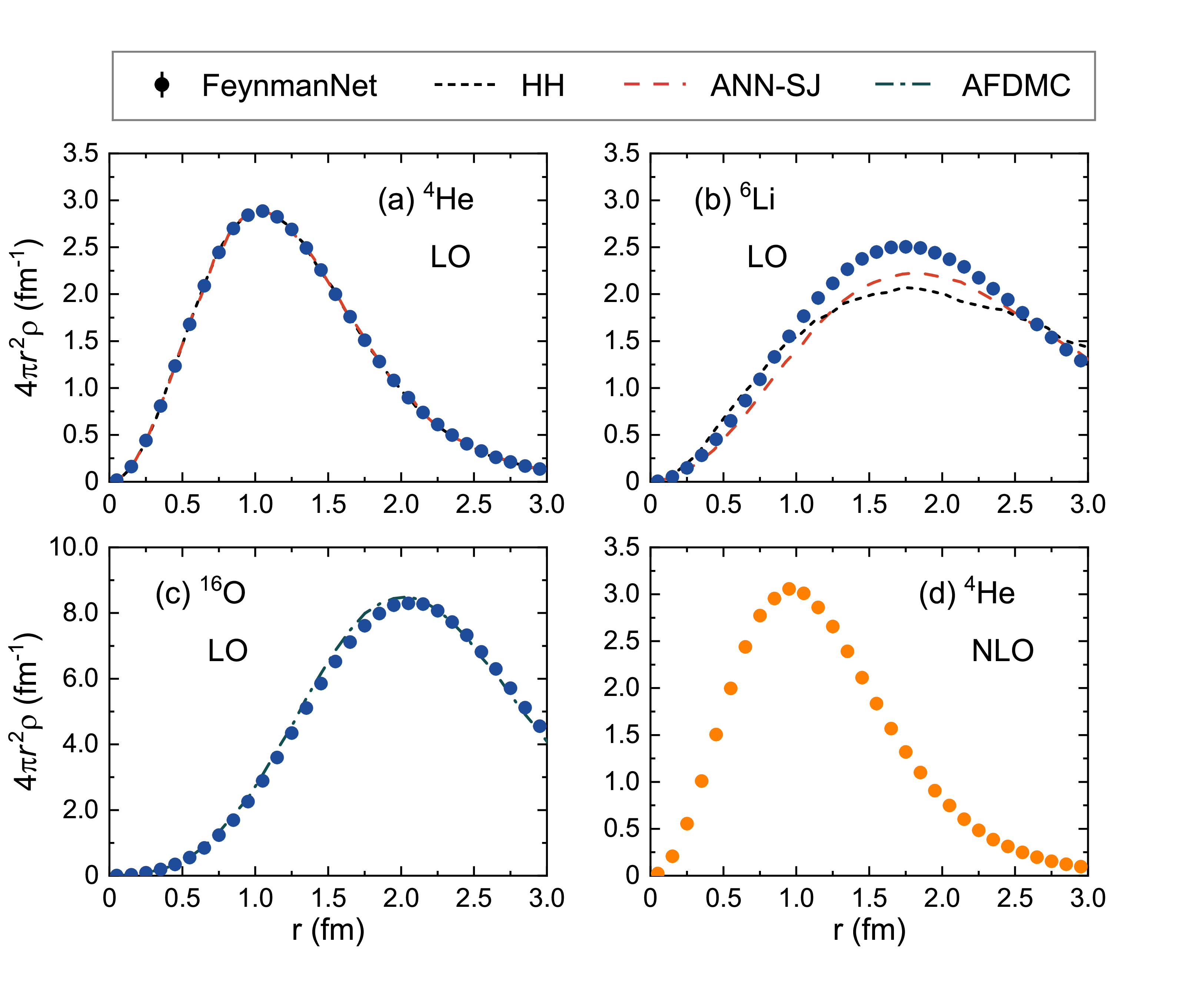}
	\caption{(Color online). Point-nucleon densities of the $^4$He, $^6$Li, and $^{16}$O ground states obtained with FeynmanNet.
The results for the LO and NLO pionless EFT Hamiltonians are shown in blue and orange, respectively.
The statistical errors from the Metropolis Monte Carlo sampling are smaller than the points.
For the LO Hamiltonian, also shown are the point-nucleon densities given by the ANN-SJ ansatz and the HH method for $^4$He and $^6$Li~\cite{Gnech2021FewBodySystems7} and the AFDMC method for $^{16}$O~\cite{Lovato2022Phys.Rev.Research043178}.}
\label{fig3}
\end{figure*}

In addition to accurate ground-state energies, FeynmanNet also provides a whole solution of the nuclear many-body wave function that, in principle, gives access to all ground-state properties.
To elucidate the quality of FeynmanNet wave function, the obtained point-nucleon densities of the $^4$He, $^6$Li, and $^{16}$O ground states are shown in Fig.~\ref{fig3}.
The point-nucleon densities are calculated as
\begin{equation}
  \rho_N(r)=\frac{1}{4\pi r^2}\frac{\langle\Psi|\sum_{i=1}^A\delta(\bar{ r}_i-r)|\Psi\rangle}{\langle\Psi|\Psi\rangle},
\end{equation}
where $\bar{r}_i$ are the distance from the $i$th nucleon to the center of mass and $\Psi$ taken to be the FeynmanNet wave function after convergence.
The obtained point-nucleon densities are compared to the previous results with the LO Hamiltonian given by the ANN-SJ ansatz and HH method for $^4$He and $^6$Li~\cite{Gnech2021FewBodySystems7} and AFDMC method for $^{16}$O~\cite{Lovato2022Phys.Rev.Research043178}.
Note that the HH method is also valid for $^4$He with the NLO Hamiltonian, but we have not found the results of point-nucleon density from the literature.

Similar to the energy, the FeynmanNet point-nucleon density for $^4$He is in an excellent agreement with the HH result (Fig.~\ref{fig3}a).
For $^6$Li, FeynmanNet provides not only the lowest ground-state energy but also a significantly more compact point-nucleon density than the ANN-SJ ansatz and the HH method (Fig.~\ref{fig3}b).
For $^{16}$O, the point-nucleon density given by FeynmanNet is very close to the AFDMC one (Fig.~\ref{fig3}c).
These results corroborate once again the accuracy of FeynmanNet in representing nuclear wave functions.
In the future, we envisage wide applications of FeynmanNet in realistic nuclear structure studies of nuclear momentum distributions, form factors, currents, etc.

\begin{figure*}[!htpb]
    \centering
	\includegraphics[width=0.9\textwidth]{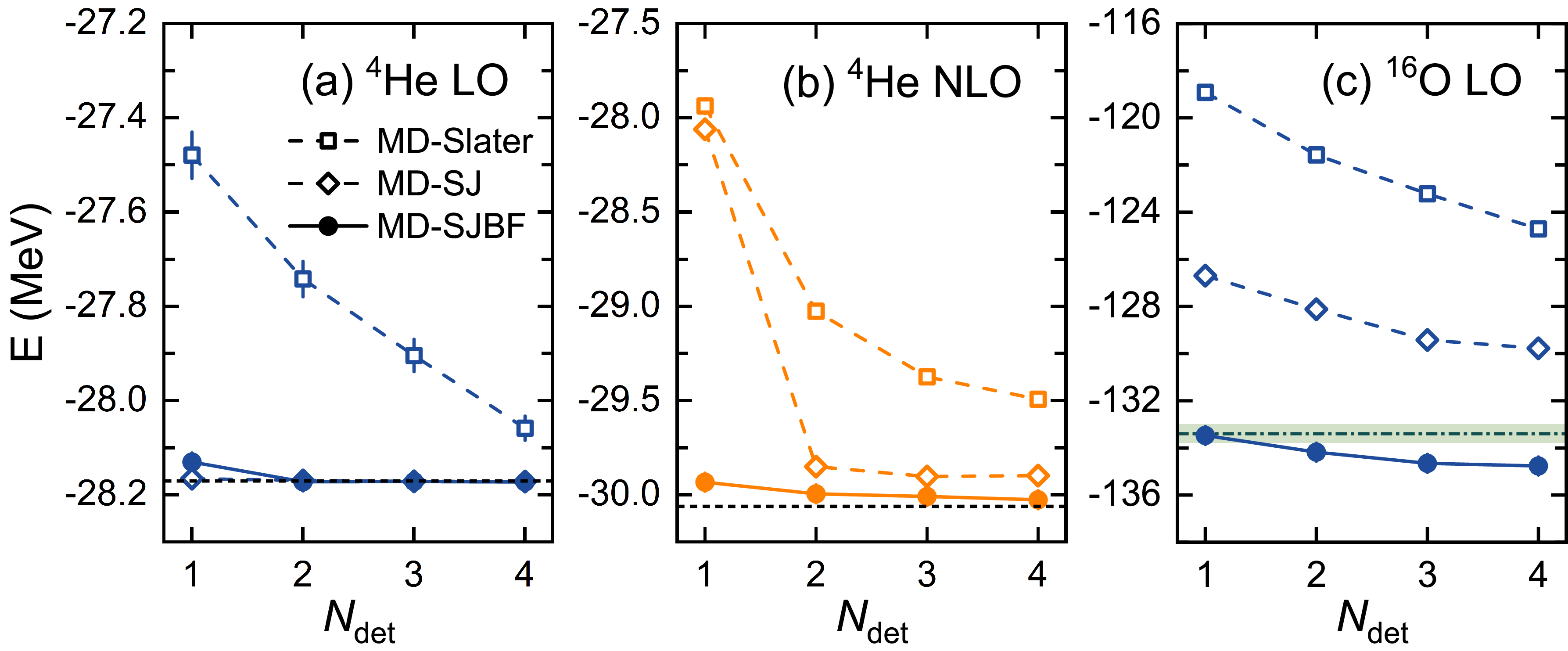}
	\caption{(Color online). Roles of the number of determinants, Jastrow factor, and backflow transformation in FeynmanNet.
The ground-state energies of $^4$He and $^{16}$O, obtained with the ansatz of multiple Slater determinants (MD) alone and in combination with the Jastrow factor (SJ) and an optional backflow transformation (BF), are shown as functions of the number of the determinants $N_\mathrm{det}$.
The numerically exact results for $^4$He from the HH method~\cite{Gnech2021FewBodySystems7} are shown by the black dashed line.
The result for $^{16}$O from the AFDMC method~\cite{Schiavilla2021Phys.Rev.C054003} is displayed by the green dash-dotted line and the shadow area indicating its statistical error.
}
\label{fig4}
\end{figure*}

Figure~\ref{fig4} highlights the specific roles of multiple Slater determinants, the Jastrow factor, and the backflow transformation in capturing nuclear many-body correlations.
We compare the $^4$He and $^{16}$O ground-state energies obtained with FeynmanNet and its two simpler variants without the backflow transformation and additionally the Jastrow factor.
For $^4$He at LO (Fig.~\ref{fig4}a), the result obtained using the ansatz of one Slater determinant alone is higher than the HH energy by about 800 keV, and this deviation can be nicely removed by the consideration of Jastrow correlations.
While both the Jastrow factor and the multiple Slater determinants are crucial to improve the energy at NLO (Fig.~\ref{fig4}b), the energy deviation from the exact value remains about 300 keV with $N_\mathrm{det}=4$, which can only be further reduced by taking into account the backflow transformation.
This should be attributed to the more complicated nodal surface of the nuclear wave function with the NLO Hamiltonian, arising from the presence of tensor and spin-orbit forces.
The Jastrow factor can compactly incorporate many-body correlations, but cannot modify the nodal surface of the Slater determinants due to its nonnegative feature.
Therefore, the backflow transformation plays a crucial role in improving the nodal surface and reaching a significantly higher accuracy.

The importance of the backflow transformation is even more evident in the larger nucleus $^{16}$O.
As seen in Fig.~\ref{fig4}c, the backflow transformation lowers the Slater-Jastrow energy by about 7 MeV and achieves, with just one determinant, an energy similar to the AFDMC result.
By increasing the number of the determinants $N_\mathrm{det}$, the calculated energy is further lowered by about 1 MeV.

\section{Summary}
We have developed FeynmanNet, a deep-learning QMC approach aiming to solve the \emph{ab initio} nuclear many-body problems, and demonstrated that it can provide very accurate solutions of $^4$He, $^6$Li, and $^{16}$O ground-state energies and wave functions emerging from the LO and NLO Hamiltonians from pionless EFT.
By introducing a well-designed backflow transformation, it outperforms the previous ANN Slater-Jastrow wave functions in not only the higher accuracy but also the nice compatibility with the Hamiltonian containing tensor and spin-orbit forces, which induce a complicated nodal surface of the wave function in the spatial and spin-isospin space.
Compared to the conventional nuclear QMC approaches, FeynmanNet has a more favorable polynomial scaling instead of an exponential scaling.
Moreover, the strong expressive power of the deep neural networks in FeynmanNet allows a variational approach to reach or even exceed the accuracy of diffusion Monte Carlo method and, thus, avoids the imaginary-time propagations that suffer from the severe inherent fermion-sign problem.
Therefore, FeynmanNet is a promising \emph{ab initio} method that can accurately solve light and medium-mass nuclei.
Note that the adopted Hamiltonians in the present work are based on the relatively simplified pionless EFT.
In the future, we plan to adopt more realistic and sophisticated chiral EFT interactions.

\begin{acknowledgments}
  This work has been supported in part by the National Key R\&D Program of China (Contract No. 2018YFA0404400),the National Natural Science Foundation of China (Grants No. 12070131001, No. 11875075, No. 11935003, No. 11975031, No. 12141501), and the High-performance Computing Platform of Peking University.
\end{acknowledgments}


%

\end{document}